\documentclass[letter,twocolumn]{jpsj3}

\usepackage[dvipdfmx]{graphicx}
\usepackage{float}
\usepackage{amsmath}
\usepackage{amssymb}
\usepackage{txfonts}
\usepackage{bm}
\usepackage{braket}
\usepackage{bxpapersize}
\usepackage{ulem}

\usepackage{color}

\newcommand{\Eq}[1]{Eq.~(\ref{#1})}
\newcommand{\Fig}[1]{Fig.~\ref{#1}}

\title{Bimeron Crystals by a Linearly Polarized AC Electric Field in Frustrated Magnets}
\author{Tatsuya Shirato$^1$, Ryota Yambe$^2$, and Satoru Hayami$^3$}
\inst{
  $^1$Department of Physics, Hokkaido University, Sapporo 060-0810, Japan\\
  $^2$Department of Applied Physics, The University of Tokyo, Tokyo 113-8656, Japan\\
  $^3$Graduate School of Science, Hokkaido University, Sapporo 060-0810, Japan}

\abst{
  We theoretically propose a method to generate topological spin textures by irradiating a classical spin system with a linearly polarized AC electric field.
  To this end, we investigate non-equilibrium steady states in a classical Heisenberg model with frustrated exchange interactions on a two-dimensional triangular lattice by numerically solving the Landau-Lifshitz-Gilbert equation at zero temperature.
  Our results reveal that the linearly polarized AC electric-field irradiation induces a topological phase transition from a single-$Q$ spiral state to a bimeron crystal with the skyrmion number of one in the low-frequency regime.
  Furthermore, we show that the obtained bimeron crystal remains relatively stable against both easy-axis and easy-plane single-ion anisotropies.
}

\begin{document}
\maketitle

Magnetic skyrmions have topologically non-trivial spin textures characterized by an integer skyrmion number $n_{\rm sk}$~\cite{skyrme1962unified, Nagaosa2013,Gobel2021}. 
They have drawn immense interest since the experimental discovery of skyrmion crystals (SkXs) in chiral magnets without the spatial inversion symmetry~\cite{Muhlbauer2009, yu2010real}.
The SkX denotes a periodic alignment of the skyrmions, which can be engineered by superpositions of multiple spiral waves with different ordering wave vectors.
Such a topologically characteristic spin configuration produces an effective magnetic field, which becomes a source of exotic physical phenomena, such as
the topological Hall effect~\cite{Neubauer2009,Jiang2016,Kurumaji2019} 
and current-driven motion~\cite{Jonietz2010,Yu2012}.

The SkXs have been so far observed in various magnets under both noncentrosymmetric and centrosymmetric lattice structures~\cite{Tokura2021}.
The materials like $\rm MnSi$~\cite{Muhlbauer2009,Jonietz2010}, $\rm FeGe$~\cite{Yu2010FeGe,Zhang2016}, and $\rm GaV_4S_8$~\cite{Kezsmarki2015,Ruff2015} are categorized into the former, while those like $\rm Gd_2PdSi_3$~\cite{Kurumaji2019,Hirschberger2020PRB, Dong_PhysRevLett.133.016401}, Gd$_3$Ru$_4$Al$_{12}$~\cite{hirschberger2019skyrmion, Hirschberger_10.1088/1367-2630/abdef9}, GdRu$_2$Si$_2$~\cite{khanh2020nanometric, khanh2022zoology, eremeev2023insight}, and $\rm GdRu_2Ge_2$~\cite{Yoshimochi2024} are categorized into the latter. 
Based on the emergence of the SkXs under various lattice structures, several stabilization mechanisms have been clarified~\cite{Tokura2021, Hayami2024MaterToday}. 
A typical mechanism is the competition between the ferromagnetic exchange interaction and the Dzyaloshinskii-Moriya interaction~\cite{Dzyaloshinsky1958, Moriya1960}, which can be applied to noncentrosymmetric lattice structures~\cite{Robler2006, Yi_PhysRevB.80.054416, Butenko_PhysRevB.82.052403}.
Meanwhile, in centrosymmetric lattice structures, alternative mechanisms have been proposed, such as frustrated exchange interactions~\cite{Okubo2012}, symmetric anisotropic interactions~\cite{amoroso2020spontaneous, yambe2021skyrmion, Yambe_PhysRevB.106.174437}, high-harmonic wave-vector interactions~\cite{hayami2022multiple}, and single-ion anisotropy~\cite{Leonov2015}.

Although important interactions to stabilize the SkXs are different between noncentrosymmetric and centrosymmetric systems, the necessity of an external magnetic field is common~\cite{Bogdanov1989, Bogdanov1994, Robler2006}.
In addition, different types of the SkXs can be induced depending on the direction of the magnetic field; a bimeron crystal with the in-plane magnetization has been found in magnets with the easy-plane magnetic anisotropy under the in-plane magnetic field~\cite{Gobel2019, Hayami2022inplane}.
In this way, the effect of the magnetic field through the Zeeman coupling is essential to induce the topological spin textures.

Meanwhile, the effect of the electric field has recently attracted attention as another way to generate an isolated skyrmion and SkXs, incorporating as
an effective coupling to multiple spins with the same symmetry as the electric field, similar to the inverse Dzyaloshinskii-Moriya mechanism~\cite{Tokura2014}.
For example, one of the multiferroic compounds, $\mathrm{Cu_2OSeO_3}$, exhibits the SkX,
where the static electric field can rotate the SkXs and change its stability region~\cite{White2012,White2014,Okamura2016} via a $d$-$p$ hybridized mechanism~\cite{Seki2012Science}.
Moreover, recent studies have shown further possibilities for nucleating the skyrmions by the electric field pulse~\cite{Mochizuki2015, Mochizuki2016, Huang2018} 
and for stabilizing the SkX by the circularly polarized electric field~\cite{Yambe2024circular}. 

\begin{figure}[t!]
  \begin{center}
  \includegraphics[width=1.0\hsize]{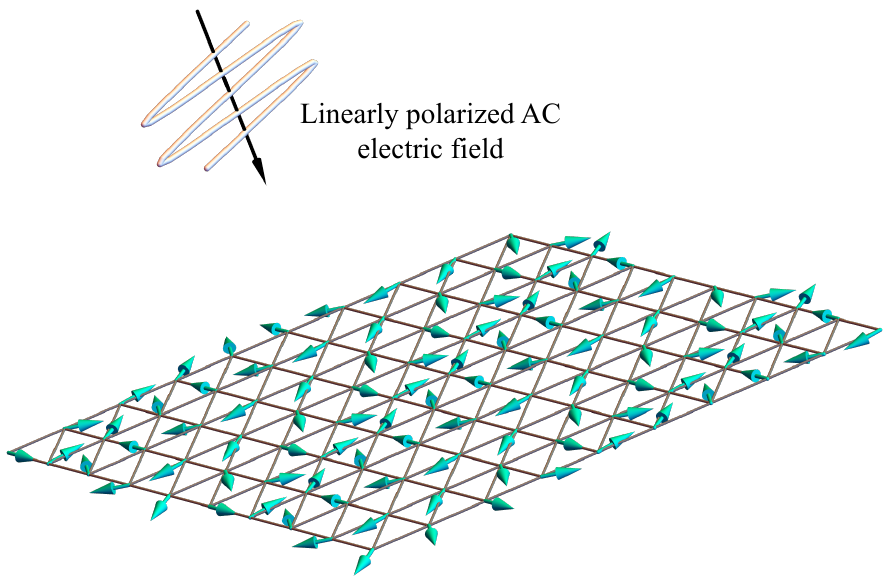} 
  \caption{\label{fig:scheme}
  (Color online). Schematic picture of the irradiation of the linearly polarized AC electric field onto the single-$Q$ spiral state with the ordering wave vector $\bm{Q}^*_1=(2\pi/5,0)$ on the triangular lattice.
  The arrows drawn by cyan represent the spin.
  }
  \end{center}
\end{figure}

In this Letter, we theoretically propose an alternative way of realizing topologically
spin ordered states using the electric field without relying on the external magnetic field.
Especially, we focus on the effect of a linearly polarized AC electric field on the stability of topological spin textures. 
We show that irradiating centrosymmetric magnets with a linearly polarized AC electric field induces
the bimeron crystal with the skyrmion number of one when the input frequency is low but not zero.
We demonstrate this by analyzing a classical Heisenberg model with frustrated competing exchange interactions on a two-dimensional triangular lattice. 
By numerically solving the Landau-Lifshitz-Gilbert (LLG) equation at zero temperature under the AC electric field, we find a topological phase transition from a single-$Q$ spiral state to a triple-$Q$ bimeron crystal in the non-equilibrium steady region. 
We also discuss the stability of the bimeron crystal when the single-ion anisotropy is introduced. 

First, we consider a classical Heisenberg model on the two-dimensional triangular lattice under a linearly polarized AC electric field.
The model Hamiltonian is given by
\begin{align}
  \mathcal{H}(t) = \mathcal{H}_{\mathrm{s}} -\bm{E}(t)\cdot\sum_{\langle i,j \rangle}\boldsymbol{p}_{ij}, \label{H}
\end{align}
\begin{align}
  \mathcal{H}_{\mathrm{s}} = -J_1\sum_{\langle i,j \rangle}\boldsymbol{S}_i\cdot\boldsymbol{S}_j
                    -J_3\sum_{\langle\langle i,j \rangle\rangle}\boldsymbol{S}_i\cdot\boldsymbol{S}_j
                    -A\sum_{i}\left(S_i^z\right)^2 \label{H_s},
\end{align}       
where 
$\bm{S}_i=(S^x_i, S^y_i, S^z_i)$ represents the localized spin at site $i$ with $|\bm{S}_i|=1$.
The first term in \Eq{H} stands for the static spin Hamiltonian without the electric field, which consists of
the ferromagnetic exchange interaction between the nearest-neighbor spins $\langle i,j \rangle$, $J_1>0$, and the antiferromagnetic exchange interaction between the third-neighbor spins $\langle\langle i,j \rangle\rangle$, $J_3<0$, as shown in \Eq{H_s}.
$\mathcal{H_{\rm s}}$ also includes the effect of single-ion anisotropy, where $A>0$ ($A<0$) corresponds to easy-axis (easy-plane) anisotropy.
The second term in \Eq{H} stands for the coupling of the electric field $\bm{E}(t)$ to the electric dipole $\bm{p}_{ij}$ between the nearest-neighbor bond;
a linearly polarized AC electric field is given by $\bm{E}(t)=E_0(0, -\sin\omega t, 0)$ with the amplitude $E_0$, the frequency $\omega$, and the periodicity $T=2\pi/\omega$, and the electric dipole is activated by the inverse Dzyaloshinskii-Moriya (spin-current) mechanism in the form of $\bm{p}_{ij} = - \lambda \bm{e}_{ij}\times(\bm{S}_i\times\bm{S}_j)$
with the magnetoelectric coupling constant $\lambda$ and the unit bond vector $\bm{e}_{ij}$~\cite{Katsura2005,Mostovoy2006,Sergienko2006,Tokura2014}.
It is noted that other external effects, such as the Zeeman coupling through the external magnetic field, are not considered for simplicity.

In the absence of the external electric field, i.e., $E_0 = 0$,
the ground state for $A=0$ is derived from the Fourier transform of the spin Hamiltonian, which is written as $-\sum_{\bm{q}}J_{\bm{q}}\bm{S_{q}}\cdot\bm{S}_{-\bm{q}}$ with wave vector $\bm{q}$.  
By setting the lattice constant to unity $a=1$, $J_{\bm{q}}$ is explicitly given by 
\begin{align}
  J_{\bm{q}}=&2J_1\left[\cos q_x +\cos\left(\frac{1}{2}q_x+\frac{\sqrt{3}}{2}q_y\right) +\cos\left(-\frac{1}{2}q_x+\frac{\sqrt{3}}{2}q_y\right)\right]\nonumber \\
  & +2J_3 \left[ \cos 2q_x +\cos\left(q_x+\sqrt{3}q_y\right) +\cos\left(-q_x+\sqrt{3}q_y\right)\right].\label{Jq}
\end{align}
The maximum $J_{\bm{q}}$ gives the ground state of the static Hamiltonian. 
For $J_1/|J_3|<4$, an incommensurate single-$Q$ spiral state with the ordering wave vector $|\bm{Q}^*|=Q^*=2\cos^{-1}[(1+\sqrt{1-2J_1/J_3})/4]$ becomes the ground state.
Due to the six-fold rotational symmetry of the triangular lattice, there are six equivalent ordering wave vectors;
$\pm\boldsymbol{Q}^*_1=\pm Q^*(1,0)$, $\pm\boldsymbol{Q}^*_2=\pm Q^*(-1/2,\sqrt{3}/2)$, and $\pm\boldsymbol{Q}^*_3=\pm Q^*(-1/2,-\sqrt{3}/2)$. 
Hereafter, we choose $J_1=1$ and $J_3=-0.5$ with $Q^*=2\pi/5$; the ground-state energy per site is given by $E_{\mathrm{gs}}=-J_1[ \cos Q^* +2\cos (Q^*/2) ]-J_3[ \cos (2Q^*) +2\cos Q^* ]$, whose absolute value is taken as the energy-scale unit.

In order to obtain non-equilibrium steady states (NESSs) in \Eq{H}, we solve the LLG equation at zero temperature by neglecting the heating effect.
The LLG equation is given by
\begin{align}
  \frac{d\bm{S}_i}{dt}=&-\frac{\gamma}{1+\alpha_{\mathrm {G}}^2}\left\{ \bm{S}_i\times\bm{B}^{\mathrm {eff}}_i(t)+ \alpha_{\mathrm {G}}\,\bm{S}_i\times[\bm{S}_i\times\bm{B}^{\mathrm {eff}}_i(t)] \right\},
  \label{LLG}
\end{align}
with the gyromagnetic ratio $\gamma$, the Gilbert damping constant $\alpha_{\mathrm {G}}$, and the effective magnetic field $\bm{B}^{\mathrm {eff}}_i(t)=-\partial\,\mathcal{H}(t)/\partial \bm{S}_i$.
We set $\gamma=1$ and $\alpha_{\mathrm{G}}=0.05$.
The LLG equation is solved by using the open software DifferentialEquations.jl.
The time scale is set as $|E_{\mathrm{gs}}|^{-1}$.
We consider the system consisting of $N=20^2$ spins under the periodic boundary conditions; we have confirmed that qualitatively similar results have been obtained for larger system sizes, e.g., $N=40^2$. 
As an initial spin configuration, we adopt the single-$Q$ spiral state, which corresponds to the ground state at $E_0 = 0$; we choose the spiral plane as the $xy$ plane, as the qualitatively similar results are obtained for the other spiral planes.

In the process of solving the LLG equation, we calculate the local scalar spin chirality~\cite{Kawamura1987,Wen1989,Kawamura1992}, which is given by 
$  \chi_{\bm{r}}(t) = \bm{S}_i(t)\cdot[ \bm{S}_j(t)\times \bm{S}_k(t)]$,
where sites $i$, $j$, and $k$ form the triangle at the position vector $\bm{r}$ in counterclockwise order.
The skyrmion number for the whole system on a discrete lattice is given by
\begin{align}
  N_{\mathrm{sk}}(t)= \frac{1}{4\pi} \sum_{\bm{r}} \Theta_{\bm{r}}(t), \label{Nsk}
\end{align}
with a skyrmion density\cite{Berg1981} $\Theta_{\bm{r}}(t)\in [-2\pi,2\pi)$;
\begin{align}
  \tan\frac{\Theta_{\bm{r}}(t)}{2} = \frac{2\bm{S}_i(t)\cdot[\bm{S}_j(t)\times \bm{S}_k(t)]}{[\bm{S}_i(t)+\bm{S}_j(t)+\bm{S}_k(t)]^2-1}.
\end{align}
The skyrmion number for the magnetic unit cell (the number of magnetic unit cells: $N_{\mathrm{unit}}$) is defined as $n_{\mathrm{sk}}(t) = N_{\mathrm{sk}}/N_{\mathrm{unit}}$.
In addition, we calculate the magnetization, which is given by
\begin{align}
M^\alpha(t) = \frac{1}{N}\sum_{i}S^\alpha_i(t) \label{mag},
\end{align} 
where $N$ is the system size and $\alpha= x,y,z$.
An in-plane magnetization is defined as 
\begin{align}
M^{xy}(t)=\sqrt{[M^{x}(t)]^2+[M^{y}(t)]^2}. \label{in-mag}
\end{align}
We also calculate momentum decomposition of spins in reciprocal space, which is given by
\begin{align}
  S_{\bm{q}}(t) = \sqrt{\frac{1}{N^2}\sum_{\alpha,i,j}S^\alpha_i(t)S^\alpha_j(t) e^{i\bm{q}\cdot(\bm{R}_i-\bm{R}_j)}},
\end{align}
where $\bm{q}$ is the wave vector and $\bm{R}_i$ is the position vector at the site $i$.
For these quantities, we compute their averages after the system reaches the NESS.
By setting the time to reach the NESS as $t_0$, the averaged quantities are given by
\begin{align}
  O &= \frac{1}{N_{\mathrm{smp}}}\sum_{n=1}^{N_{\mathrm{smp}}} O(t_0+n\Delta)  \label{average},
\end{align}
with the number of samples $N_{\mathrm{smp}}$ and the time step $\Delta$.
The typical values of $t_0$, $N_{\mathrm{smp}}$, and $\Delta$ are given by $t_0=800T$, $N_{\mathrm{smp}}=20000$, and $\Delta=0.02T$, respectively.

\begin{figure}[t!]
  \begin{center}
  \includegraphics[width=1.0\hsize]{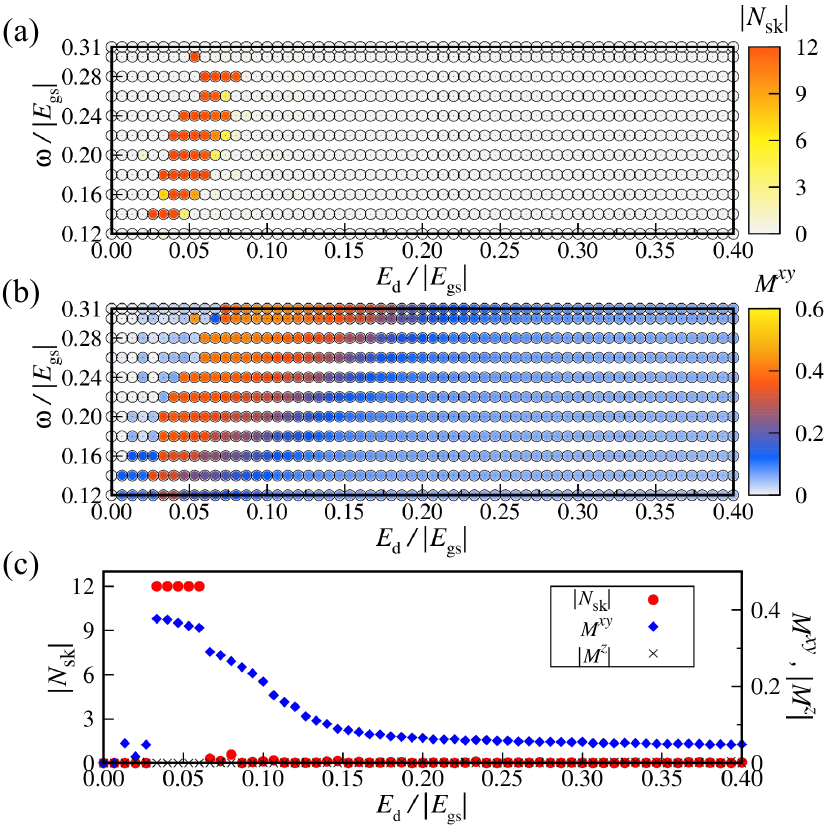} 
  \caption{\label{fig:PD1}(Color online).
  Absolute values of (a) averaged skyrmion number $N_{\rm sk}$ and (b) averaged in-plane magnetization $M^{xy}$ as functions of $E_{\rm d}$ and $\omega$ at $A=0$ in the NESSs.
  (c) $E_{\rm d}$ dependence of $N_{\rm sk}$, $M^{xy}$, and averaged out-of-plane magnetization $M^{z}$ at $\omega/|E_{\rm gs}|=0.18$ for $A=0$. }
  \end{center}
\end{figure}

\begin{figure*}[t!]
  \begin{center}
  \includegraphics[width=1.0\hsize]{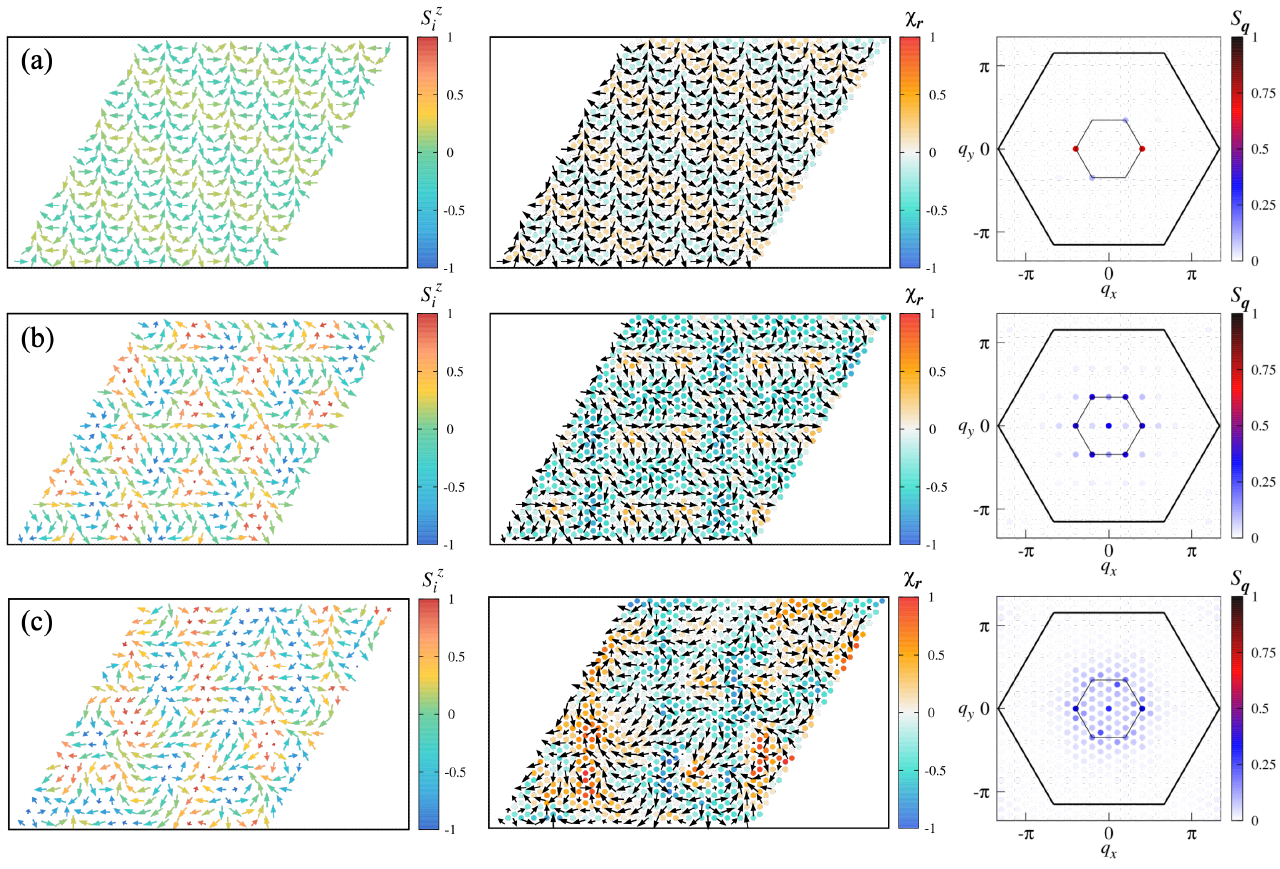} 
  \caption{\label{fig:conf}(Color online).
  Snapshots of the NESSs with $\omega/|E_{\rm gs}|=0.18$ at $A=0$;
  (a) $E_{\rm d}/|E_{\rm gs}|=0.006$, (b) $E_{\rm d}/|E_{\rm gs}|=0.033$, and (c) $E_{\rm d}/|E_{\rm gs}|=0.067$.
  (Left panels) Real-space spin configurations; the arrows and color denote the spins $\bm{S}_i$ and the $z$ component $S^z_i$, respectively.
  (Middle panels) Local scalar spin chirality configurations; the arrows and color of the circles denote the spins $\bm{S}_i$ and the local scalar spin chirality $\chi_{\bm{r}}$, respectively.
  (Right panels) Reciprocal-space spins $S_{\bm{q}}$.
  The large hexagons represent the first Brillouin zone.
  The vertices of the small hexagons correspond to $\pm\boldsymbol{Q}^*_1$, $\pm\boldsymbol{Q}^*_2$, and $\pm\boldsymbol{Q}^*_3$.
  }
  \end{center}
\end{figure*}

Figures~\ref{fig:PD1}(a) and \ref{fig:PD1}(b) show the $E_{\rm d}\,(=\lambda E_0)$ and $\omega$ dependence at $A=0$ of the time-averaged skyrmion number $|N_{\rm sk}|$ and the time-averaged in-plane magnetization $M^{xy}$, respectively, which are calculated for the NESS. 
As shown in Fig.~\ref{fig:PD1}(a), the topological state with nonzero skyrmion number appears in the low-frequency regime of $0.14 \lesssim \omega/|E_{\rm gs}| \lesssim 0.3$, which indicates that the linearly polarized AC electric field generates a topological spin texture. 
Such a state is stabilized only for the relatively small effective electric field $E_{\rm d}$. 
For example, in the case of $\omega/|E_{\rm gs}|=0.18$, the region for $0.03 \lesssim E_{\rm d}/|E_{\rm gs}| \lesssim 0.06$ 
exhibits $|N_{\rm sk}|=12$, while the other regions show no skyrmion numbers, as shown in \Fig{fig:PD1}(c). 
As another characteristic point, this state with $|N_{\rm sk}|=12$ shows the spontaneous in-plane magneization $M^{xy}$ rather than the out-of-plane one $M^z$, as shown in Figs.~\ref{fig:PD1}(b) and \ref{fig:PD1}(c).

Next, let us discuss the change of the spin configurations against $E_{\rm d}$ by choosing $\omega/|E_{\rm gs}|=0.18$; the following discussions hold for the other frequency at the qualitative level. 
For $E_{\rm d}=0$, the single-$Q$ spiral state with $\bm{Q}_1$ is realized. 
By introducing $E_{\rm d}$, the single-$Q$ spiral spin configuration is modulated so as to have the additional modulation at $\bm{Q}_3$, as shown in the right panel of Fig.~\ref{fig:conf}(a).
In other words, the single-$Q$ spiral state changes into the anisotropic double-$Q$ state for $E_{\rm d} \neq 0$. 
The additional modulation at $\bm{Q}_3$ affects the $z$-spin configuration, as shown in the left panel of Fig.~\ref{fig:conf}(a), which indicates that the double-$Q$ spin configuration is noncoplanar. 
Indeed, the density wave in terms of the scalar spin chirality appears along the $\bm{Q}_3$, as shown in the middle panel of Fig.~\ref{fig:conf}(a)~\cite{Hayami_PhysRevB.94.174420}. 
In this double-$Q$ state, the magnetizations $M^{xy}$ and $M^z$ as well as the skyrmion number $N_{\rm sk}$ almost remain zero, as shown in Fig.~\ref{fig:PD1}(c).

The double-$Q$ state turns into the state with $|N_{\rm sk}|=12$ at around $E_{\rm d} \simeq 0.033$. 
As shown by the real-space spin configuration in the left panel of Fig.~\ref{fig:conf}(b), this state consists of the periodic alignment of vortices around $S_i^z = -1$ and the anti-vortices around $S_i^z = +1$. 
Since the vortex and anti-vortex have the positive and negative winding numbers, respectively, both of them exhibit a meron spin texture with the negative skyrmion number of $-0.5$. 
In other words, a pair of vortex and anti-vortex leads to the skyrmion number of $-1$, which indicates the appearance of the bimeron~\cite{Gobel2019, Hayami2022inplane}; the number of $|N_{\rm sk}|=12$ means the twelve bimerons in the whole system so that the bimeron crystal is formed. 
Indeed, the negative scalar spin chirality distribution is found in the middle panel of Fig.~\ref{fig:conf}(b). 
It is noted that the energies between the states with positive and negative scalar spin chirality are degenerate; such a degeneracy is lifted when the bond-dependent-type anisotropy is taken into account~\cite{Hayami_PhysRevB.103.054422}. 
This bimeron crystal is characterized by the triple-$Q$ superposition, as shown in the right panel of Fig.~\ref{fig:conf}(b), although the peak intensities at $\bm{Q}^*_1$--$\bm{Q}^*_3$ are slightly different since the applied electric field breaks the threefold rotational symmetry.

With a further increase of $E_{\rm d}$, the bimeron crystal becomes unstable, which changes into another spin state without $N_{\rm sk}$. 
We show the real-space spin configuration at $E_{\rm d}/|E_{\rm gs}|=0.067$ in the left panel of Fig.~\ref{fig:conf}(c); although the vortex and anti-vortex textures are locally found, they are disordered in the whole system without $N_{\rm sk}$. 
The disordered feature is also found in the scalar spin chirality distribution in the middle panel of Fig.~\ref{fig:conf}(c). 
In addition, the peak structure in the reciprocal-space spins is smeared out, as shown in the right panel of Fig.~\ref{fig:conf}(c). 
Such a melting behavior for large $E_{\rm d}$ was also found in the case of the irradiation of the circularly polarized electric field~\cite{Yambe2024circular}. 
It is noted that the net in-plane magnetization also vanishes in this state for large $E_{\rm d} \gtrsim 0.06$. 
The nonzero values of $M^{xy}$ in Fig.~\ref{fig:PD1}(c) are due to the finite-size effect; we confirmed that $M^{xy}$ for $E_{\rm d} \gtrsim 0.06$ tends to decrease and approach zero for larger system sizes, while that for the bimeron crystal is almost unchanged.

We discuss the stabilization mechanism of the bimeron crystal. 
In contrast to previous studies~\cite{Gobel2019, Hayami2022inplane}, the present bimeron crystal under the linearly polarized AC electric field requires neither the Dzyaloshinskii-Moriya interaction nor the easy-plane magnetic anisotropy. 
In addition, the mechanism is also different from an effective electric-field-induced three-spin chiral interaction, since such an effective interaction appears only for the circularly polarized electric field in the high-frequency regime~\cite{Yambe2023}. 
Moreover, in contrast to the bimeron crystal induced by the circularly polarized electric field, the sign of $N_{\rm sk}$ is undetermined by the external electric field~\cite{Yambe2024circular}.
Thus, the present mechanism based on the linearly polarized AC electric field provides another root to realize the bimeron crystal. 

Finally, we investigate the effect of single-ion magnetic anisotropy on the bimeron crystal by setting $A\neq0$ in \Eq{H_s}, which often affects the stability region of topological spin textures~\cite{Leonov2015, Hayami_PhysRevB.99.094420}.
We fix the frequency at $\omega/|E_{\mathrm{gs}}|=0.18$, where we use the same energy unit $E_{\mathrm{gs}}$ when $A=0$.
Owing to the magnetic anisotropy, the spiral plane of the single-$Q$ spiral state realized for $E_{\rm d}=0$ is fixed on the $xy$ ($yz$) plane for $A<0$ ($A>0$); we set such a single-$Q$ spiral state as the initial spin configuration in the LLG simulations.

\begin{figure}[t!]
  \begin{center}
  \includegraphics[width=1.0\hsize]{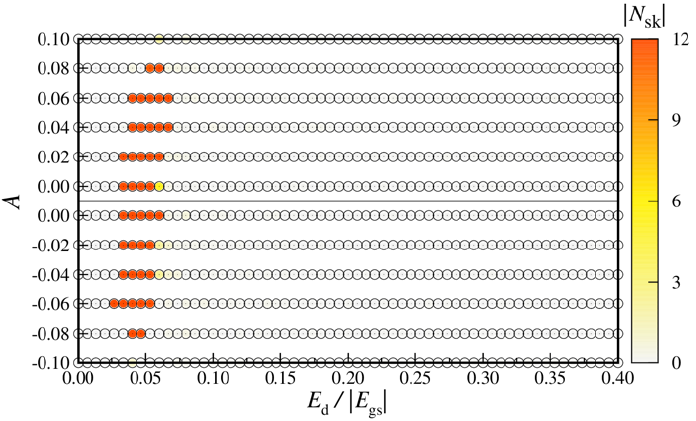} 
  \caption{\label{fig:PD2}(Color online).
  Averaged skyrmion number $|N_{\rm sk}|$ as functions of $E_{\rm d}$ and $A$ at $\omega/|E_{\mathrm{gs}}|=0.18$ in the NESSs.
  Of the two results at $A=0$, the upper (lower) one corresponds to the NESS when the initial spin configuraiton is chosen as the single-$Q$ spiral state lying on the $xy$ ($yz$) plane.
  }
  \end{center}
\end{figure}

Figure~\ref{fig:PD2} shows the $E_{\rm d}$ and $A$ dependence of the time-averaged skyrmion number $|N_{\rm sk}|$.
For $A=0$, we show the results when the initial spin configuration is the single-$Q$ spiral state lying on the $xy$ and $yz$ planes.
As shown in Fig.~\ref{fig:PD2}, the bimeron crystal remains stable for both easy-axis $(A>0)$ and easy-plane $(A<0)$ anisotropies up to $|A|\simeq 0.08$; the region of the bimeron crystal against $E_{\rm d}$ is almost unchanged irrespective of $A$.
When $|A|$ increases, the bimeron crystal changes into other topologically trivial magnetic states without $N_{\rm sk}$.
This result indicates that the bimeron crystal is expected to be realized when the single-ion magnetic anisotropy is relatively small.

To summarize, we have theoretically demonstrated that topological spin textures can be generated in frustrated multiferroic materials by applying the linearly polarized AC electric field without the magnetic field.
By applying the linearly polarized AC electric field into the single-$Q$ spiral state in the classical frustrated Heisenberg model on the triangular lattice and performing the LLG simulations at zero temperature, we have obtained the bimeron crystal with the skyrmion number of one in the low-frequency regime.
We have also investigated the stability of the bimeron crystal against the single-ion anisotropy, and we have shown that the bimeron crystal remains stable for both easy-axis and easy-plane single-ion anisotropies.

The present results provide the important conditions to generate the bimeron crystal by the linearly polarized AC electric field; the frustrated exchange interaction leading to the single-$Q$ spiral state without the external field and the relatively small magnetic anisotropy.
Given these conditions, a compound $\mathrm{NiX_2}$ $(\mathrm{X = Br, I})$ is one of the potential candidate materials, 
as it can form a single-layer triangular lattice with competing exchange interactions~\cite{Song2022,Prayitno2024}. 
Especially, previous studies have reported that $\mathrm{NiI_2}$ shows multiferroic properties~\cite{Song2022,Liu2024}.
Our findings would stimulate an experimental realization of electric-field-induced SkXs in such multiferroic materials with frustrated exchange interactions.

\begin{acknowledgment}
  T. S. is grateful to Y. Zha for fruitful discussions.
  This research was supported by JSPS KAKENHI Grants Numbers JP21H01037, JP22H00101, JP22H01183, JP23H04869, JP23K03288, JP23K20827, and by JST CREST (JPMJCR23O4) and JST FOREST (JPMJFR2366).
\end{acknowledgment}

\bibliographystyle{jpsj}
\bibliography{jpsj_letter2}

\begin{thebibliography}{10}

\bibitem{skyrme1962unified}
T.~H.~R. Skyrme, Nucl. Phys. {\bfseries 31},  556 (1962).

\bibitem{Nagaosa2013}
N.~Nagaosa and Y.~Tokura, Nat. Nanotech. {\bfseries 8},  899 (2013).

\bibitem{Gobel2021}
B.~G^^c3^^b6bel, I.~Mertig, and O.~A. Tretiakov, Phys. Rep. {\bfseries 895},  1 (2021).

\bibitem{Muhlbauer2009}
S.~M^^c3^^bchlbauer, B.~Binz, F.~Jonietz, C.~Pfleiderer, A.~Rosch, A.~Neubauer, R.~Georgii, and P.~B^^c3^^b6ni, Science {\bfseries 323},  915 (2009).

\bibitem{yu2010real}
X.~Z. Yu, Y.~Onose, N.~Kanazawa, J.~H. Park, J.~H. Han, Y.~Matsui, N.~Nagaosa, and Y.~Tokura, Nature {\bfseries 465},  901 (2010).

\bibitem{Neubauer2009}
A.~Neubauer, C.~Pfleiderer, B.~Binz, A.~Rosch, R.~Ritz, P.~G. Niklowitz, and P.~B^^c3^^b6ni, Phys. Rev. Lett. {\bfseries 102},  186602 (2009).

\bibitem{Jiang2016}
W.~Jiang, X.~Zhang, G.~Yu, W.~Zhang, X.~Wang, M.~B. Jungfleisch, J.~E. Pearson, X.~Cheng, O.~Heinonen, K.~L. Wang, Y.~Zhou, A.~Hoffmann, and S.~G.~T. Velthuis, Nat. Phys. {\bfseries 13},  162 (2016).

\bibitem{Kurumaji2019}
T.~Kurumaji, T.~Nakajima, M.~Hirschberger, A.~Kikkawa, Y.~Yamasaki, H.~Sagayama, H.~Nakao, Y.~Taguchi, T.~hisa Arima, and Y.~Tokura, Science {\bfseries 365},  914 (2019).

\bibitem{Jonietz2010}
F.~Jonietz, S.~M^^c3^^bchlbauer, C.~Pfleiderer, A.~Neubauer, W.~M^^c3^^bcnzer, A.~Bauer, T.~Adams, R.~Georgii, P.~B^^c3^^b6ni, R.~A. Duine, K.~Everschor, M.~Garst, and A.~Rosch, Science {\bfseries 330},  1648 (2010).

\bibitem{Yu2012}
X.~Z. Yu, N.~Kanazawa, W.~Z. Zhang, T.~Nagai, T.~Hara, K.~Kimoto, Y.~Matsui, Y.~Onose, and Y.~Tokura, Nat. Commun. {\bfseries 3},  988 (2012).

\bibitem{Tokura2021}
Y.~Tokura and N.~Kanazawa, Chem. Rev. {\bfseries 121},  2857 (2021).

\bibitem{Yu2010FeGe}
X.~Z. Yu, N.~Kanazawa, Y.~Onose, K.~Kimoto, W.~Z. Zhang, S.~Ishiwata, Y.~Matsui, and Y.~Tokura, Nat. Mater. {\bfseries 10},  106 (2010).

\bibitem{Zhang2016}
L.~Zhang, H.~Han, M.~Ge, H.~Du, C.~Jin, W.~Wei, J.~Fan, C.~Zhang, L.~Pi, and Y.~Zhang, Sci. Rep. {\bfseries 6},  22397 (2016).

\bibitem{Kezsmarki2015}
I.~K^^c3^^a9zsm^^c3^^a1rki, S.~Bord^^c3^^a1cs, P.~Milde, E.~Neuber, L.~M. Eng, J.~S. White, H.~M. R^^c3^^b8nnow, C.~D. Dewhurst, M.~Mochizuki, K.~Yanai, H.~Nakamura, D.~Ehlers, V.~Tsurkan, and A.~Loidl, Nat. Mater. {\bfseries 14},  1116 (2015).

\bibitem{Ruff2015}
E.~Ruff, S.~Widmann, P.~Lunkenheimer, V.~Tsurkan, S.~Bord^^c3^^a1cs, I.~K^^c3^^a9zsm^^c3^^a1rki, and A.~Loidl, Science Advances {\bfseries 1},  1500916 (2015).

\bibitem{Hirschberger2020PRB}
M.~Hirschberger, T.~Nakajima, M.~Kriener, T.~Kurumaji, L.~Spitz, S.~Gao, A.~Kikkawa, Y.~Yamasaki, H.~Sagayama, H.~Nakao, S.~Ohira-Kawamura, Y.~Taguchi, T.-h. Arima, and Y.~Tokura, Phys. Rev. B {\bfseries 101},  220401 (2020).

\bibitem{Dong_PhysRevLett.133.016401}
Y.~Dong, Y.~Arai, K.~Kuroda, M.~Ochi, N.~Tanaka, Y.~Wan, M.~D. Watson, T.~K. Kim, C.~Cacho, M.~Hashimoto, D.~Lu, Y.~Aoki, T.~D. Matsuda, and T.~Kondo, Phys. Rev. Lett. {\bfseries 133},  016401 (2024).

\bibitem{hirschberger2019skyrmion}
M.~Hirschberger, T.~Nakajima, S.~Gao, L.~Peng, A.~Kikkawa, T.~Kurumaji, M.~Kriener, Y.~Yamasaki, H.~Sagayama, H.~Nakao, K.~Ohishi, K.~Kakurai, Y.~Taguchi, X.~Yu, T.-h. Arima, and Y.~Tokura, Nat. Commun. {\bfseries 10},  5831 (2019).

\bibitem{Hirschberger_10.1088/1367-2630/abdef9}
M.~Hirschberger, S.~Hayami, and Y.~Tokura, New J. Phys. {\bfseries 23},  023039 (2021).

\bibitem{khanh2020nanometric}
N.~D. Khanh, T.~Nakajima, X.~Yu, S.~Gao, K.~Shibata, M.~Hirschberger, Y.~Yamasaki, H.~Sagayama, H.~Nakao, L.~Peng, K.~Nakajima, R.~Takagi, T.-h. Arima, Y.~Tokura, and S.~Seki, Nat. Nanotechnol. {\bfseries 15},  444 (2020).

\bibitem{khanh2022zoology}
N.~D. Khanh, T.~Nakajima, S.~Hayami, S.~Gao, Y.~Yamasaki, H.~Sagayama, H.~Nakao, R.~Takagi, Y.~Motome, Y.~Tokura, T.-h. Arima, and S.~Seki, Adv. Sci. {\bfseries 9},  2105452 (2022).

\bibitem{eremeev2023insight}
S.~Eremeev, D.~Glazkova, G.~Poelchen, A.~Kraiker, K.~Ali, A.~V. Tarasov, S.~Schulz, K.~Kliemt, E.~V. Chulkov, V.~Stolyarov, et~al., Nanoscale Adv. {\bfseries 5},  6678 (2023).

\bibitem{Yoshimochi2024}
H.~Yoshimochi, R.~Takagi, J.~Ju, N.~D. Khanh, H.~Saito, H.~Sagayama, H.~Nakao, S.~Itoh, Y.~Tokura, T.~Arima, S.~Hayami, T.~Nakajima, and S.~Seki, Nat. Phys. {\bfseries 20},  1001 (2024).

\bibitem{Hayami2024MaterToday}
S.~Hayami and R.~Yambe, Mater. Today Quantum {\bfseries 3},  100010 (2024).

\bibitem{Dzyaloshinsky1958}
I.~Dzyaloshinsky, J. Phys. Chem. Solids {\bfseries 4},  241 (1958).

\bibitem{Moriya1960}
T.~Moriya, Phys. Rev. {\bfseries 120},  91 (1960).

\bibitem{Robler2006}
U.~K. R^^c3^^b6^^c3^^9fler, A.~N. Bogdanov, and C.~Pfleiderer, Nature {\bfseries 442},  797 (2006).

\bibitem{Yi_PhysRevB.80.054416}
S.~D. Yi, S.~Onoda, N.~Nagaosa, and J.~H. Han, Phys. Rev. B {\bfseries 80},  054416 (2009).

\bibitem{Butenko_PhysRevB.82.052403}
A.~B. Butenko, A.~A. Leonov, U.~K. R\"o\ss{}ler, and A.~N. Bogdanov, Phys. Rev. B {\bfseries 82},  052403 (2010).

\bibitem{Okubo2012}
T.~Okubo, S.~Chung, and H.~Kawamura, Phys. Rev. Lett. {\bfseries 108},  017206 (2012).

\bibitem{amoroso2020spontaneous}
D.~Amoroso, P.~Barone, and S.~Picozzi, Nat. Commun. {\bfseries 11},  5784 (2020).

\bibitem{yambe2021skyrmion}
R.~Yambe and S.~Hayami, Sci. Rep. {\bfseries 11},  11184 (2021).

\bibitem{Yambe_PhysRevB.106.174437}
R.~Yambe and S.~Hayami, Phys. Rev. B {\bfseries 106},  174437 (2022).

\bibitem{hayami2022multiple}
S.~Hayami, J. Phys. Soc. Jpn. {\bfseries 91},  023705 (2022).

\bibitem{Leonov2015}
A.~O. Leonov and M.~Mostovoy, Nat. Commun. {\bfseries 6},  8275 (2015).

\bibitem{Bogdanov1989}
A.~N. Bogdanov and D.~A. Yablonskii, Sov. Phys. JETP {\bfseries 68},  101 (1989).

\bibitem{Bogdanov1994}
A.~Bogdanov and A.~Hubert, J. Magn. Magn. Mater. {\bfseries 138},  255 (1994).

\bibitem{Gobel2019}
B.~G^^c3^^b6bel, A.~Mook, J.~Henk, I.~Mertig, and O.~A. Tretiakov, Phys. Rev. B {\bfseries 99},  060407(R) (2019).

\bibitem{Hayami2022inplane}
S.~Hayami, Phys. Rev. B {\bfseries 103},  224418 (2021).

\bibitem{Tokura2014}
Y.~Tokura, S.~Seki, and N.~Nagaosa, Rep. Prog. Phys. {\bfseries 77},  076501 (2014).

\bibitem{White2012}
J.~S. White, I.~Levati^^c4^^87, A.~A. Omrani, N.~Egetenmeyer, K.~Pr^^c5^^a1a, I.~^^c5^^bdivkovi^^c4^^87, J.~L. Gavilano, J.~Kohlbrecher, M.~Bartkowiak, H.~Berger, and H.~M. R^^c3^^b8nnow, J. Phys.: Condens. Matter {\bfseries 24},  432201 (2012).

\bibitem{White2014}
J.~S. White, K.~Pr^^c5^^a1a, P.~Huang, A.~A. Omrani, I.~^^c5^^bdivkovi^^c4^^87, M.~Bartkowiak, H.~Berger, A.~Magrez, J.~L. Gavilano, G.~Nagy, J.~Zang, and H.~M. R^^c3^^b8nnow, Phys. Rev. Lett. {\bfseries 113},  107203 (2014).

\bibitem{Okamura2016}
Y.~Okamura, F.~Kagawa, S.~Seki, and Y.~Tokura, Nat. Commun. {\bfseries 7},  1 (2016).

\bibitem{Seki2012Science}
S.~Seki, X.~Z. Yu, S.~Ishiwata, and Y.~Tokura, Science {\bfseries 336},  198 (2012).

\bibitem{Mochizuki2015}
M.~Mochizuki and Y.~Watanabe, Appl. Phys. Lett. {\bfseries 107},  082409 (2015).

\bibitem{Mochizuki2016}
M.~Mochizuki, Adv. Electron. Mater. {\bfseries 2},  1500180 (2016).

\bibitem{Huang2018}
P.~Huang, M.~Cantoni, A.~Kruchkov, J.~Rajeswari, A.~Magrez, F.~Carbone, and H.~M. R^^c3^^b8nnow, Nano Lett. {\bfseries 18},  5167 (2018).

\bibitem{Yambe2024circular}
R.~Yambe and S.~Hayami, Phys. Rev. B {\bfseries 110},  014428 (2024).

\bibitem{Katsura2005}
H.~Katsura, N.~Nagaosa, and A.~V. Balatsky, Phys. Rev. Lett. {\bfseries 95},  057205 (2005).

\bibitem{Mostovoy2006}
M.~Mostovoy, Phys. Rev. Lett. {\bfseries 96},  067601 (2006).

\bibitem{Sergienko2006}
I.~A. Sergienko and E.~Dagotto, Phys. Rev. B {\bfseries 73},  094434 (2006).

\bibitem{Kawamura1987}
H.~Kawamura and M.~Tanemura, Phys. Rev. B {\bfseries 36},  7177 (1987).

\bibitem{Wen1989}
X.~G. Wen, F.~Wilczek, and A.~Zee, Phys. Rev. B {\bfseries 39},  11413 (1989).

\bibitem{Kawamura1992}
H.~Kawamura, Phys. Rev. Lett. {\bfseries 68},  3785 (1992).

\bibitem{Berg1981}
B.~Berg and M.~L^^c3^^bcscher, Nucl. Phys. B {\bfseries 190},  412 (1981).

\bibitem{Hayami_PhysRevB.94.174420}
S.~Hayami, S.-Z. Lin, Y.~Kamiya, and C.~D. Batista, Phys. Rev. B {\bfseries 94},  174420 (2016).

\bibitem{Hayami_PhysRevB.103.054422}
S.~Hayami and Y.~Motome, Phys. Rev. B {\bfseries 103},  054422 (2021).

\bibitem{Yambe2023}
R.~Yambe and S.~Hayami, Phys. Rev. B {\bfseries 108},  064420 (2023).

\bibitem{Hayami_PhysRevB.99.094420}
S.~Hayami and Y.~Motome, Phys. Rev. B {\bfseries 99},  094420 (2019).

\bibitem{Song2022}
Q.~Song, C.~A. Occhialini, E.~Erge^^c3^^a7en, B.~Ilyas, D.~Amoroso, P.~Barone, J.~Kapeghian, K.~Watanabe, T.~Taniguchi, A.~S. Botana, S.~Picozzi, N.~Gedik, and R.~Comin, Nature {\bfseries 602},  601 (2022).

\bibitem{Prayitno2024}
T.~B. Prayitno, E.~Budi, and Y.~P. Sarwono, J. Phys. Chem. Solids {\bfseries 188},  111910 (2024).

\bibitem{Liu2024}
N.~Liu, C.~Wang, C.~Yan, C.~Xu, J.~Hu, Y.~Zhang, and W.~Ji, Phys. Rev. B {\bfseries 109},  195422 (2024).

\end{thebibliography}

\end{document}